\documentclass[twocolumn,prx,aps,superscriptaddress]{revtex4}
\usepackage{amssymb,amsmath,amsfonts}

\usepackage[latin9]{inputenc}
\usepackage{amstext}
\usepackage{mathtools}

\DeclarePairedDelimiter\ket{\lvert}{\rangle}
\DeclarePairedDelimiterX\braket[2]{\langle}{\rangle}{#1\,\delimsize\vert\,\mathopen{}#2}
\usepackage{enumitem}
\usepackage{graphicx,bm,palatino}
\usepackage[colorlinks=true,linkcolor=blue,urlcolor=blue,citecolor=blue,pdfusetitle]{hyperref}

\usepackage[sc]{mathpazo} 

\usepackage{hyperref,cleveref}
\usepackage[dvipsnames]{xcolor}
\usepackage[caption=false]{subfig}

\usepackage{times}
\usepackage{bbm}

\makeatother

\begin{document}
\title{
Characterizing the spontaneous collapse of a wavefunction through entropy production}
\date{\today}

\author{Simone Artini (email: \href{mailto:simone.artini@you.unipa.it}{simone.artini@you.unipa.it})}
\affiliation{Department of Physics, University of Trieste, Strada Costiera 11, 34151 Trieste, Italy}
\author{Mauro Paternostro (email: \href{mailto:m.paternostro@qub.ac.uk}{m.paternostro@qub.ac.uk})}
\affiliation{Universit\`a degli Studi di Palermo, Dipartimento di Fisica e Chimica - Emilio Segr\`e, via Archirafi 36, I-90123 Palermo, Italy}
\affiliation{Centre for Quantum Materials and Technologies, School of Mathematics and Physics, Queen's University Belfast, BT7 1NN, United Kingdom}

\begin{abstract}
We investigate the phenomenology leading to the non-conservation of energy of the continuous spontaneous localization (CSL) model from the viewpoint of non-equilibrium thermodynamics, and use such framework to assess the equilibration process entailed by the dissipative formulation of the model (dCSL). As a paradigmatic situation currently addressed in frontier experiments aimed at investigating possible collapse theories, we consider a one-dimensional mechanical oscillator in a thermal state. We perform our analysis in the phase space of the oscillator,
where the entropy production rate, a non-equilibrium quantity used to characterize irreversibility, can be conveniently analyzed.  
We show that the CSL model violates Clausius law, as it exhibits a negative entropy production rate, while the dCSL model reaches equilibrium consistently only under certain dynamical conditions, thus allowing us to identify the values -- in the parameter space -- where the latter mechanism can be faithfully used to describe a thermodynamically consistent phenomenon.
\end{abstract}

\maketitle

\section{Introduction}
\label{sec:intro}

The quantum-to-classical transition, which is the process driving the quantum state of a system towards a fully classical description of its physical configuration, is yet to achieve a full characterization and, most remarkably, the satisfactory understanding of its underlying mechanism~\cite{Bell}. Particularly relevant is the question on whether the loss of coherence experienced by a large and complex quantum system should be ascribed to an intrinsic mechanism or the unavoidable presence of the surrounding environment~\cite{BassiRMP}. As environmental decoherence only provides a partial addressing of the measurement problem, alternative theoretical frameworks, where the collapse of the wavefunction is lifted to the rank of a universal mechanism, are currently being formulated and developed to attack the quantum-to-classical transition~\cite{bassi2003dynamical, carlesso2022present,Carlesso2018,Carlesso2019}. Such {\it collapse models}   are achieved through stochastic dynamics, and are usually characterized by phenomenological parameters that are asked to satisfy criteria of reasonability based, for instance, on the retrieval of a classical description in the macroscopic limit. 
The Continuous Spontaneous Localisation (CSL), one of the most well-studied of such models~\cite{bassi2003dynamical}, describes the loss of coherence in the position basis by
way of an an extra dissipative term entering the master equation of a quantum system~\cite{ghirardi1986unified,ghirardi1990markov,pearle1989combining}.  The physical intuition behind it is that the wave function of the physical state of a system undergoes random localization processes, called ``jumps", occurring due to a dissipative mechanism not ascribable
to any of the other environmental noise source, and taking place at a rate that depends on the dimension of the system itself: while microscopic systems are left basically unaffected, linear superpositions of states of a macroscopic system would be strongly suppressed due to an intrinsic amplification mechanism. Mathematically, this is achieved by interpreting the wavefunction as a stochastic process in the Hilbert space~\cite{ghirardi1990markov}.  

Despite its apparent simplicity and appeal, the CSL model suffers of the fundamental shortfall of being inherent not energy-conserving: albeit at a very slow rate, the expectation value of the energy of a quantum system of mass $m$ undergoing CSL-like dynamics grows indefinitely with time, thus signalling the fundamental unphysical nature of the model. A dissipative extension of this model --  dubbed the dCSL model --- has been proposed~\cite{smirne2014dissipative}, which, while still not conserving the energy, introduces a term that acts as friction, allowing  energy to reach an asymptotic finite value and thus an effective temperature at which the system thermalizes. 

In this paper, we provide an original characterization of CSL and dCSL model from the perspective offered by non-equilibrium thermodynamics: by using a phase-space formulation of irreversible entropy~\cite{santos2017wigner, santos2018irreversibility}, which aptly quantifies the degree of thermodynamic irreversibility of a given physical process~\cite{landipaternostroRMP}, we address the collapse-affected dynamics of a quantum harmonic oscillator subjected to either CSL or dCSL. We show that, while the standard CSL model implies the violation of Clausius law of thermodynamics, witnessed by the occurrence of negative entropy production rates, the dCSL extension would result in thermodynamically consistent descriptions, under suitably chosen dynamical conditions, despite the explicit lack of energy conservation, thus embodying a more plausible formulation of a collapse mechanism to consider. In providing such an assessment, we identify regimes of the dCSL model where, despite a dominant dissipative character of the dynamics, a violation of the Second Law of thermodynamics is enabled by suitably squeezing the initial state of the oscillator. 
 By addressing the features of fundamental collapse theories from a completely general thermodynamic standing point, our work demonstrates the intrinsic value of non-equilibrium tools for the characterization of open quantum system dynamics.

The remainder of this paper is organized as follows. After briefly reviewing the salient features of CSL and dCSL models [cf. Sec.~\ref{modelli}], in Sec.~\ref{sec:FP}, the corresponding quantum Fokker-Planck equations are solved numerically for an initial thermal state of the oscillator.  In Sec.~\ref{sec:ent}, the  quantities used in our thermodynamic analysis will be briefly introduced, and the entropy production rate calculated numerically for the case study and the results discussed in the manuscript. Finally, relevant concluding remarks are offered in Sec.~\ref{sec:conc}, while a technical Appendix reports details of the calculations required for the phase-space formulation of the dynamics.  

\section{CSL and dCSL model: an introduction}
\label{modelli}

The stochastic differential equation that describes the evolution of a state under the action of a collapse mechanism such as one of those at the centre of our study 
is $d\ket{\psi(t)}=\hat{\cal O}\ket{\psi(t)}$, where we have introduced the operator $\hat{\cal O}$ that acts on a generic state vector $\ket{a}$ as
\begin{equation}
\label{modsch}
\hat{\cal O}\!\ket{a}{=}\left[-\frac{i\hat H}{\hbar}
{+}\gamma\int d^3{x}\hat N({\bf x})dB({\bf x}) {-}\frac{\gamma}{2}\int d^3x\hat N^2({\bf x})dt\right]\!\ket{a}.
\end{equation}
Here, $B(\bf x)$ is a continuous set of Wiener processes and $\hat{N}({\bf x}) = \sum_{s}\int d^3{\bf  y} g({\bf y}-{\bf x}) \hat{a}^{\dag}({\bf y},s)\hat{a}({\bf y},s)$ is an averaged number operator -- $ \hat{a}$ and $ \hat{a}^{\dag}$ being the annihilation and creation operators of a harmonic oscillator -- written in the second quantization formalism, with a Gaussian weighing function $g\left(\bf x\right) = \bigl(\frac{\alpha}{2\pi}\bigr)^{\frac{3}{2}}e^{-\frac{\alpha}{2}(\bf x)^2}\>$. Such a Gaussian weight is an assumption of the model that defines the 
length at which the suppression of macroscopic linear superpositions takes place, as will be discussed shortly.  In this equation, two important parameters are present: the intensity of the Markovian noise entailed by the Wiener process  $\gamma$, which is related to the rate of the jumps, and the length-scale $\alpha$, which is related to the typical localization volume $V_{loc}= {\alpha ^{-3/2}}$. The master equation of the CSL for the statistical operator can be shown to be
\begin{equation}
\label{mastereqcsl}
\begin{aligned}
\frac{d\hat{\rho}(t)}{dt} &= -\frac{i}{\hbar}\left[\hat{H},\hat{\rho}(t)\right] +\gamma \int d^3 x \hat{N}\left(\bf x\right)\hat{\rho}\left(t\right)\hat{N}\left(\bf x\right)  \\ 
& -\frac{\gamma}{2}\int d^3 x \left\{ \hat{N}^2\left(\bf x\right),  \hat{\rho}(t)\right\}\> .
\end{aligned}
\end{equation}

Some important results can be derived. Firstly, the off-diagonal elements of the statistical operator in the position basis go to zero exponentially fast when considering distances greater than the typical localization length ${1}/{\sqrt{\alpha}}$, which, together with the fact that the localization happens at the wave function level, guarantees the effective suppression of macroscopic linear superpositions. The choice of the parameters $\gamma$ and $\alpha$ can be made in such a way that the localization happens on very short time scales for objects made of a large number of particles (i.e. of the order of the Avogadro's number), while leaving the standard Schr\"odinger evolution for systems made of few particles basically uneffected (the aforementioned amplification mechanism). Furthermore, the expectation values of the position and of the momentum evolve in time like in the unitary evolution and the Ehrenfest theorem holds and the internal degrees of freedom are decoupled from the center of mass as in the standard quantum theory. However in this simple formulation the energy is not conserved and it is, on the contrary, divergent in time as
\begin{equation}
\langle \hat{H} \rangle = \langle \hat{H} \rangle_{Sch} + \frac{\lambda\alpha\hbar^2}{4m}t \>,
\label{energiacsl}
\end{equation}
where $\langle \cdot \rangle$ denotes the quantum expectation value taken with respect to the modified dynamic for $\hat{\rho}$, while $\langle \cdot \rangle_{Sch}$ is the quantum expectation value taken with respect to the standard Schr\"odinger dynamic \cite{bassi2003dynamical}. The dCSL extension of this model~\cite{smirne2014dissipative}, fails to conserve energy but introduces in Eq.~\eqref{modsch} a term that depends on the momentum through a new parameter in the weighing function that acts as friction, thus allowing the energy to reach an asymptotic finite value and thus an effective temperature at which the system thermalizes.

\section{Numerical solution of the dynamics in the phase-space picture}
\label{sec:FP}
\subsection{The quantum Fokker-Planck equation}

In order to carry out a thermodynamical analysis of the dynamics, it is necessary to translate the master equation, concerning the statistical operator, into a Fokker-Planck equation written in terms of the Wigner function \cite{navarretebenlloch2022introduction} of the system. The reason behind this will be clear in Sec. \ref{sec:ent} as it lies on the choice of the entropy that will be used and, furthermore, allows for a simple numerical solution of the case study. This is achieved, as far as the CSL model is concerned, via the Wigner-Weyl transform \cite{manko2002alternative} of eq. \eqref{mastereqcsl}. The full computation is carried out in the Appendix. One can show that, for a one-dimensional system and considering gaussian states, i.e. states with gaussian Wigner function (as the thermal state of the oscillator, for example), the Fokker-Planck equation of the system can be well approximated with
\begin{equation}
\partial_t W_{\hat{\rho}}(q,p) = \bigl\{ W_{\hat{H}}, W_{\hat{\rho}} \bigr\}_*(q,p) +D\,\partial^2_{p}W_{\hat{\rho}}(q,p),
\label{fp2}
\end{equation}
where:  $W_{\hat{H}}$ is the Weyl symbol of the Hamiltonian, $W_{\hat{\rho}}$ is the Weyl symbol of the statistical operator, i.e. the Wigner function of the system, and $\bigl\{ W_{\hat{H}}, W_{\hat{\rho}} \bigr\}_*$ is the Moyal bracket \cite{manko2002alternative} of the two symbols that comes from the unitary term. This is written explicitly \cite{baker1958formulation,manko2002alternative} as
\begin{equation}
\label{moyaldef}
\bigl\{ W_{\hat{H}}, W_{\hat{\rho}} \bigr\}_* = 2 W_{\hat{H}}\sin\left(\frac{1}{2}\left[(\overleftarrow{\partial_q},\overrightarrow{\partial_p}\right] \right) W_{\hat{\rho}} ,
\end{equation}
where we have omitted the arguments of the Wigner function for simplicity of notation. Here, $D = \sqrt{{\gamma ^2 \alpha ^3}/{\pi}}$ is the diffusion parameter of the dynamics. Indeed, eq. \eqref{fp2} shows that the simplified collapse term is a simple anisotropic diffusion in the momentum direction. In what follows, unless otherwise specified, we use natural units according to which $\hbar = 1$ and rescale the position and momentum operators as ${p}/{p_{zpf}} \to p$ and ${q}/{q_{zpf}} \to q$ with $p_{zpf}=\sqrt{{mw}/{2}}$ and $q_{zpf}={{1}/\sqrt{2mw}}$ the zero-point fluctuations of an harmonic oscillator with mass $m$ and frequency $w$. Correspondingly, we have ${\alpha}/{2mw} \rightarrow \alpha$. Considering instead the dCSL model, it has been shown in Ref.~\cite{smirne2014dissipative} that the modified Schr\"odinger equation leads to an asymptotic value of the mean energy
\begin{equation}
\langle \hat{H} \rangle = (\langle \hat{H} \rangle_{Sch} - \langle \hat{H} \rangle_{as})e^{-\xi t}  + \langle \hat{H} \rangle_{as}, 
\end{equation}
with $\langle \hat{H} \rangle_{as}={\hbar^2\alpha}/(16mk)$, $\xi = \dfrac{\gamma( {\alpha}/{\pi})^{{3}/{2}}}{2(1+k)^2}$ and $k$ is related to the parameter introduced in the momentum dependent term in the modified Schr\"odinger equation. The equilibrium temperature of the system 
can thus be written as $T={\alpha \hbar^2}/{(8k_B k)}$ and it is estimated to be $T\simeq 10^{-1}K$. The friction effect driving the system to such equilibrium configuration can be accounted for in the Fokker-Planck equation by adding a dissipative term to Eq.~\eqref{fp2} as follows
\begin{equation}
\label{fpdcsl}
\partial_t W_{\hat{\rho}} = \left\{ W_{\hat{H}}, W_{\hat{\rho}} \right\}_* +D\,\partial^2_{p}W_{\hat{\rho}} + \partial_p\left( f p W_{\hat{\rho}}\right),
\end{equation}
where $f$ is the dissipative constant and natural units and dimensionless $q$, $p$ are considered once again.

\subsection{CSL model: numerical solution of the dynamics in phase space}
\begin{figure} [b!]
    \centering
    \includegraphics[width=0.8\columnwidth]{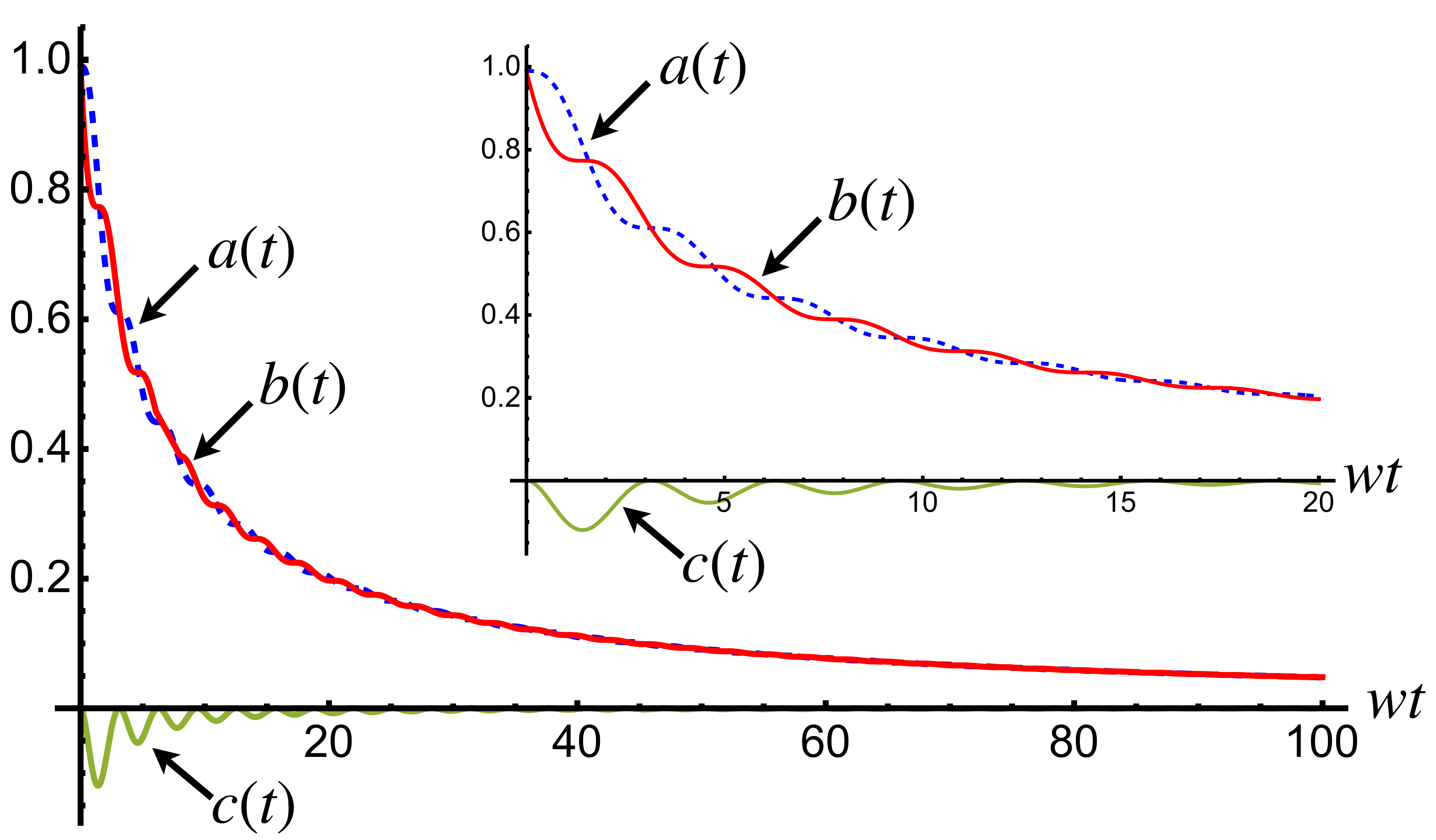}
    \caption{Evolution of $a(t)$ (dashed blue line), $b(t)$ (solid red line), and $c(t)$ (solid green curve) against time. All quantities are dimensionless. In this simulation we have used the parameters {$D/(m\omega)=0.1$}, ${a_0}={b_0}=1/1.01$, $c_0 = 0$. The inset shows the behavior in a shorter time-window to appreciate the nearly out-of-phase nature of $a(t)$ and $b(t)$ instigated by the uncertainty principle. 
    }
    \label{fig1}
\end{figure}

Keeping the natural units, but restoring the proper dimensions of the phase-space variables, we consider the initial Wigner function
\begin{equation}
W_{\hat{\rho}}(q,p) = \dfrac{a_0}{\pi}\exp\left[ - a_0 \biggl(  mw q^2 + \dfrac{p^2}{mw} \biggl) \right]
\end{equation}
and the generic ansatz
\begin{equation}
W_{\hat{\rho}}(q,p) = \frac{\sqrt{a(t)b(t) - c^2(t)}}{\pi}e^{ - \bigl(a(t) mwq^2 + b(t) \frac{p^2}{mw} + 2c(t)pq \bigl)}
\end{equation}
to describe the anisotropic evolved state at a generic time of the dynamics. Here, the dimensionless time-dependent parameters $a(t)$, $b(t)$ and $c(t)$ need to be determined from the evolution of the system. Let us call $\Sigma=V$ the covariance matrix of the system, where $V ^{-1}=\begin{pmatrix}a&c\\c&b\end{pmatrix}$. 
We start by looking into  the unitary term: from Eq.~\eqref{moyaldef} it is straightforward to check that the Moyal bracket is equal to the Poisson bracket up to order $\hbar$, that is $\bigl\{ W_{\hat{H}}, W_{\hat{\rho}} \bigr\}_* = \bigl\{ W_{\hat{H}}, W_{\hat{\rho}} \bigr\} + {\cal O}(\hbar^2)$. 
Furthermore, one can show that 
$W_{\hat{H}}(q,p)= ({p^2}/{m} + {mw^2}q^2)/2$~\cite{navarretebenlloch2022introduction}. By using the ansatz in the Fokker-Planck equation and equating the terms with the same powers of $p$, $q$ and $pq$, we get the following set of differential equations 
\begin{equation}
\label{second_pair}
\dot{a} = 2wc -\frac{4Dc^2}{mw}, \dot{b} = -2wc -\frac{4Db^2}{mw}, \dot{c} = w(b{-}a) -\frac{4Dbc}{mw}
\end{equation}
with the additional condition 
$\frac{d}{dt}\ln(ab-c^2)
= -\frac{4Db}{mw}$.
While these equations do not admit a stationary solution, it is straightforward to gather the temporal behavior of $a(t)$ and $b(t)$. 
We take $D/(m w)=0.1$ as diffusion coefficient, ${b_0}={a_0}=1/1.01$ and $c_0 = 0$ as initial conditions and integrating over the dimensionless evolution time $\omega t$, thus finding the behavior illustrated in Fig.~\ref{fig1}.
%
Clearly, the dominant effect is diffusion, leading to a progressive spread of the Wigner function that only reaches a non-equilibrium steady state. On the other hand the unitary term causes a rotation in phase-space which is responsible for the emergence of transient correlations. This will cause the variances to fluctuate around the linear increasing trend of the diffusion, which would not be present in the $q$ direction without the unitary term. Notice that without the diffusion term, the rotation in the phase-space would not affect a symmetric  Wigner function such as that of a thermal state, which is in fact the stationary solution of the unitary dynamics. 

\subsection{Dissipative CSL model: numerical solution of the dynamics in phase space}

\begin{figure} [t!]
    \centering
{\bf (a)}\\
    \includegraphics[width=0.8\columnwidth]{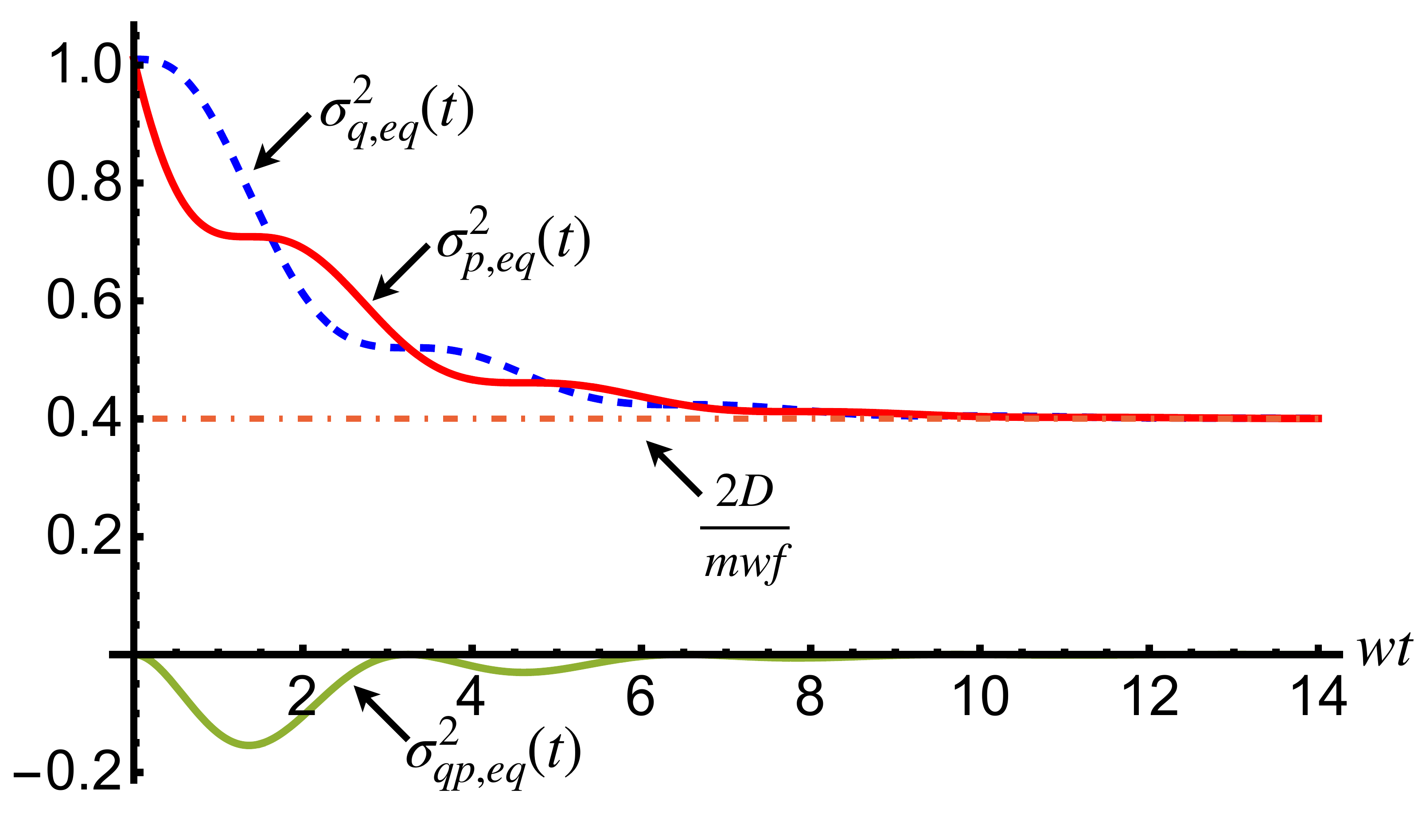}\\  
    {\bf (b)}\\
   \includegraphics[width=0.8\columnwidth]{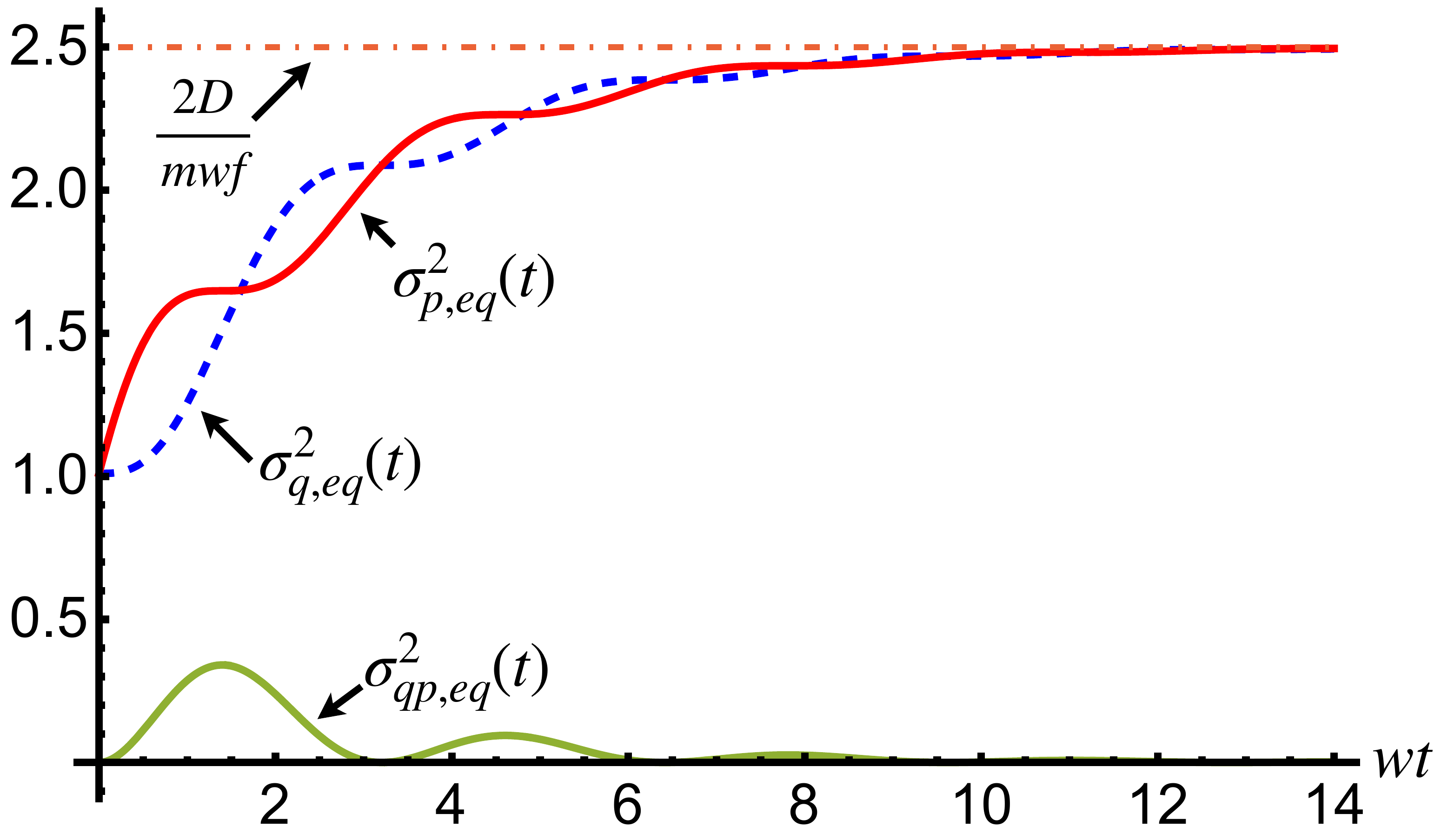}
    \caption{Evolution of the variance of  position (dashed blue line) and momentum (solid red line), and of the covariance $\Sigma_{12}(t)$ for the dCSL model. In panel {\bf (a)} we look into the dynamics under dominant-friction conditions by taking $D/(mw)=0.5$, $f=0.7$. Panel {\bf (b)} is  diffusion-dominated as we have chosen $D/(mw)=0.5$, $f=0.4$. The initial conditions are ${a_0}={b_0}=1/1.01$, $c_0 = 0$ in both panels. The quantities being plotted are all dimensionless. }
    \label{dominantfrictiondiffusion}
\end{figure}

The approach sketched above can be used also for the dCSL model, finding the dynamical equations
\begin{equation}
\label{second_pair2}
\begin{aligned}
&\dot{a} = 2wc -\frac{4Dc^2}{mw} \> , \\
& \dot{b} = -2wc -\frac{4Db^2}{mw} +2fb, \\
&\dot{c} = w(b -a) -\frac{4Dbc}{mw} +fc\> 
\end{aligned}
\end{equation}
with the further constraint $\frac{d}{dt}\ln(ab-c^2)=-4D b/(mw)+2f$. The additional, $f$-dependent terms in the dynamical equations lead to a non-trivial isotropic stationary solution characterized by the equilibrium parameters $c_{eq} = 0$ and $a_{eq} = b_{eq} = {mwf}/{(2D)}$. 
Such isotropic state can be seen as a thermal state with a finite effective temperature $T$ determined by the following relation {
\begin{equation}
a_{eq} = \dfrac{1}{2}\dfrac{e^{\frac{w}{T}}-1}{e^{\frac{w}{T}}+1}=\dfrac{mw}{2D}f.
\end{equation}}
Depending on the relative value of $D/mw$ and $f$, two cases can be identified: For the diffusion-dominated case where $f<{D}/{mw}$, the final variances will end up being larger than any initial value. The phenomenology is the opposite in the friction-dominated case corresponding to the choice $f>{D}/mw$. This is well illustrated in Fig.~\ref{dominantfrictiondiffusion}, where we show the convergence of the elements of the covariance matrix of the system to the asymptotic values. 


\begin{figure} [t!]
    \centering
    \includegraphics[width=0.49\linewidth]{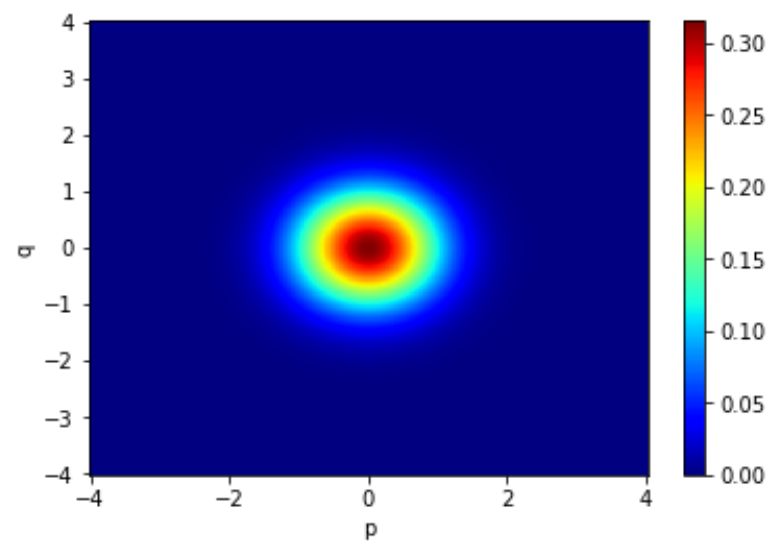}
    \includegraphics[width=0.49\linewidth]{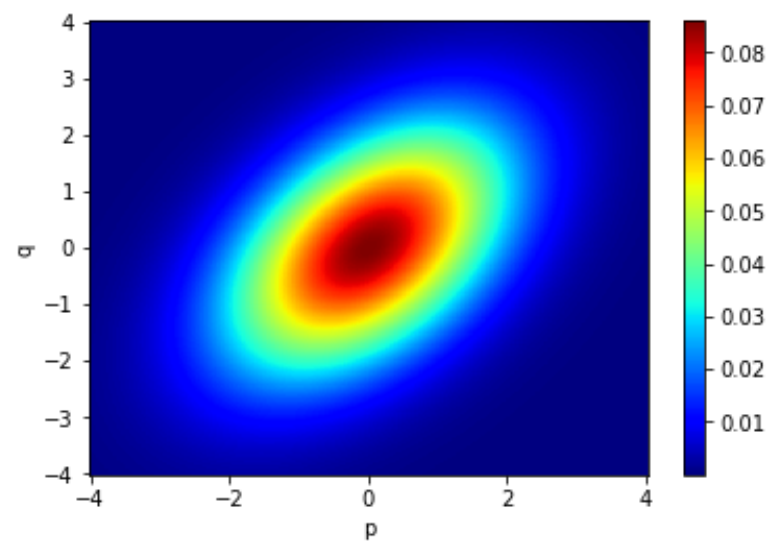}
    \centering
    \includegraphics[width=0.49\linewidth]{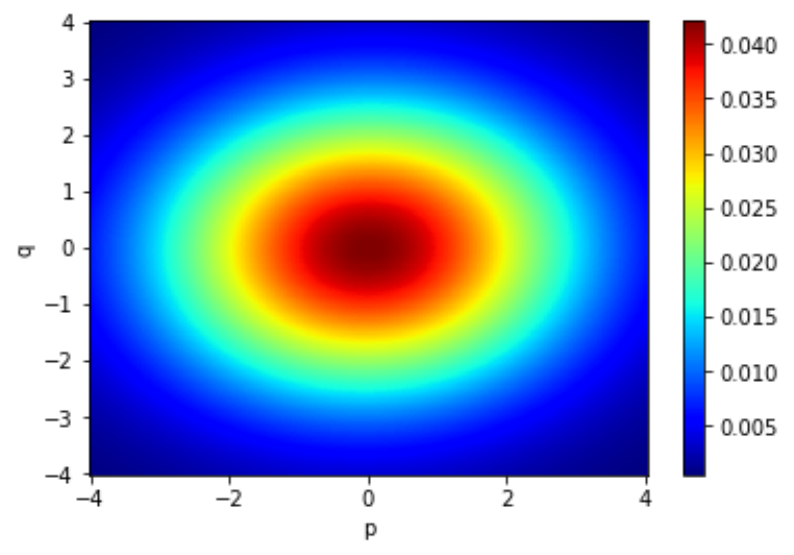}
    \includegraphics[width=0.49\linewidth]{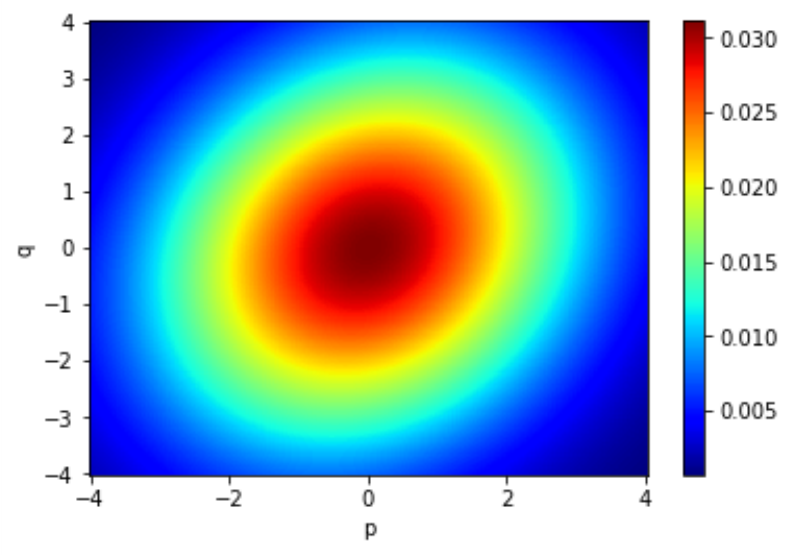}
    \caption{Snapshots of the dynamics of the Wigner function of the system under the effects of the CSL mechanism. We sample the distribution at four different times ($wt=0,1.6,3.2,5$). All quantities are dimensionless. In this simulation we have used the parameters $D/(m\omega)=0.9$, ${a_0}={b_0}=1/1.01$, $c_0 = 0$.}
    \label{csl_timestamps}
\end{figure}

\begin{figure} [b!]
    \centering
    \includegraphics[width=0.49\linewidth]{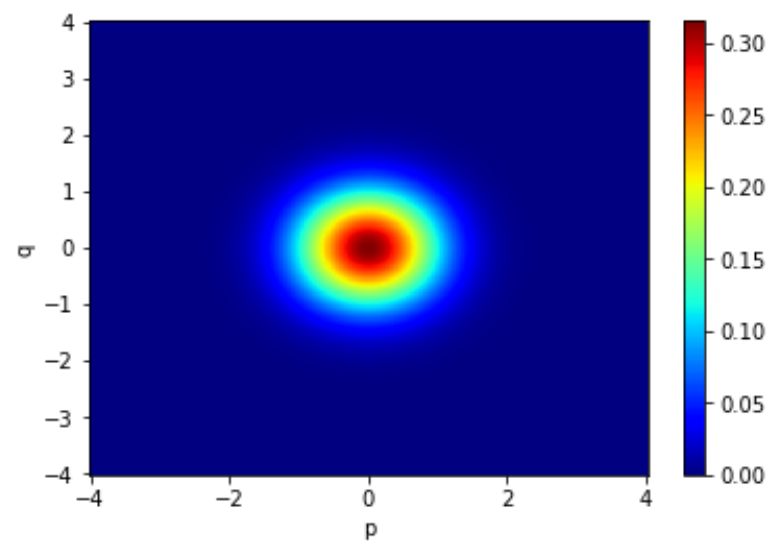}
    \includegraphics[width=0.49\linewidth]{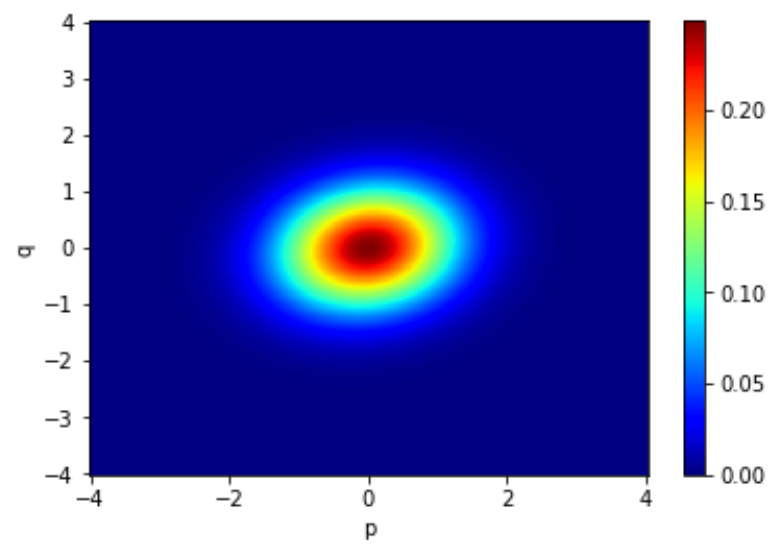}
    \centering
    \includegraphics[width=0.49\linewidth]{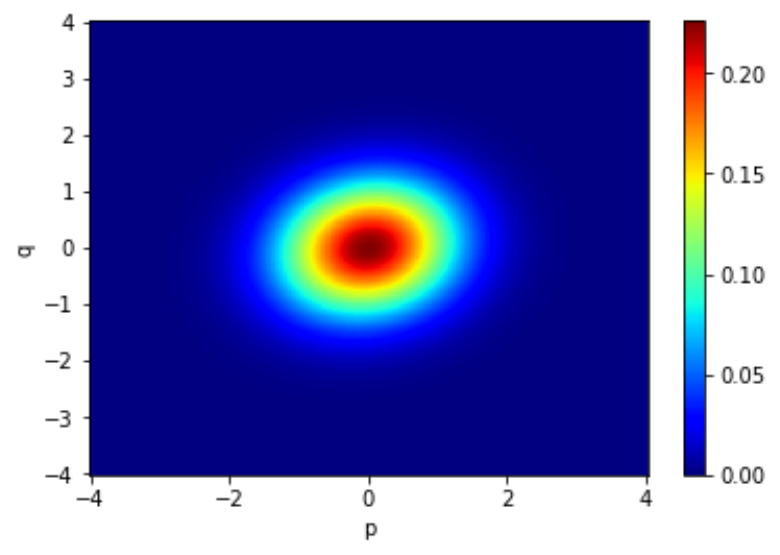}
    \includegraphics[width=0.49\linewidth]{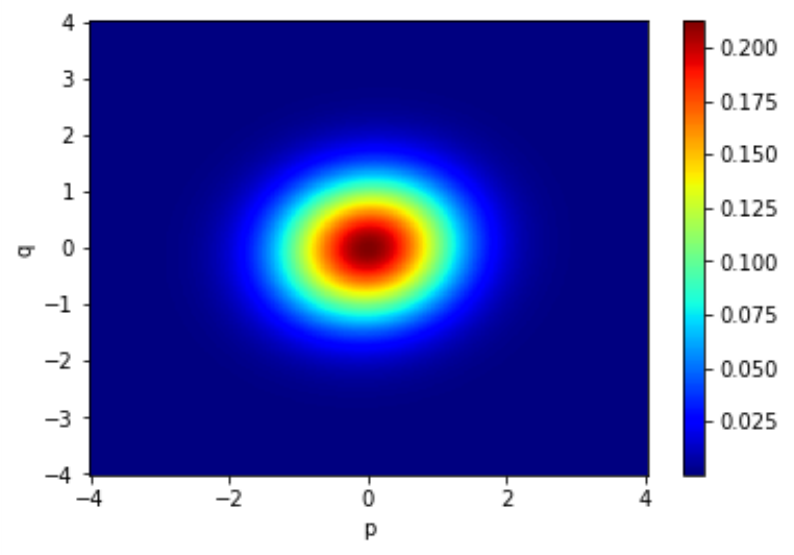}
    \caption{napshots of the dynamics of the Wigner function of the system under the effects of the dCSL mechanism. We sample the distribution at four different times($wt=0,32,68,98$). All quantities are dimensionless. In this simulation we have used the parameters $D/(m\omega)=1.6$, $f=2$, ${a_0}={b_0}=1/1.01$, $c_0 = 0$.}
    \label{dcsl_timestamps}
\end{figure}

As in the non-dissipative dynamics, the distribution is streched along the  $p$ direction by the diffusion and rotated by the unitary term. This time however the dissipation competes with the friction
until the distribution settles around a symmetric state whose variances, should  the diffusion term dominate, would be larger than the initial values [cf. Fig.~\ref{csl_timestamps} and Fig.~\ref{dcsl_timestamps} for a qualitative comparison between the CSL and the dCSL dynamics of the Wigner function of the system].

\section{Entropic analysis of the collapse dynamics}
\label{sec:ent}
\subsection{The definition of entropy and the entropy production rate}
Having characterized the phase-space dynamics of the system under the collapse models at the centre of this study, we now introduce the thermodynamical quantities used in the present paper. The main theoretical tool is the entropy production~\cite{landipaternostroRMP,velasco2011entropy}, i.e. the contribution to the total entropy of a thermodynamic transformation or process that is produced by taking or keeping the system out of equilibrium. 
It embodies a quantitative measure of irreversibility in such processes and its rate is used to determine whether a system approaches thermal equilibrium during its dynamics ~\cite{deffner2011nonequilibrium}.
For a general open-system dynamics, the entropy production rate $\Pi$ is defined as~\cite{landipaternostroRMP}
\begin{equation}
\dfrac{dS}{dt} = \Pi(t) - \phi (t),
\end{equation} 
where $\phi(t)$ is the entropy flux between the system and the environment it is in contact with. 
Its thermodynamically consistent definition implies the request for the entropy production rate to satisfy a class of fluctuation theorems, namely mathematical generalizations of the second law of thermodynamics~\cite{seifert2008stochastic}, thus giving rise to the constraint $\Pi(t)\ge0$ across a dynamics.

When working in the phase space, a successful formulation of the framework for the quantification of entropy production, which allows to bypass some of the shortfalls of the standard approach based on the use of von Neumann entropy (such as the so-called {\it ultra-cold catastrophe}~\cite{uzdinCatastrophe}), makes use of the R\'enyi-$2$ entropy~\cite{santos2017wigner, santos2018irreversibility} defined as $S_{2}=-\mathrm{ln}(\mathrm{Tr}{\hat\rho}^{2})$, where $\hat\rho$ is the density matrix of the system. For Gaussian states, such quantity is equivalently formulated --  up to an irrelevant constant -- as 
\begin{equation}
S_2(t) {=}\frac12\ln[\det V(t)]{=} -\int W(q, p,t)\ln\bigl( W(q, p,t) \bigr) dq dp
\label{wigent}
\end{equation}
when expressed in terms of the covariance matrix $V(t)$ and Wigner function $W(p,q,t)$ at time $t$ associated with $\hat\rho$~\cite{adesso2012measuring}. 
With such tool at hand, the entropy production rate is then defined as~\cite{santos2017wigner,landipaternostroRMP}
\begin{equation}
\label{PI}
\Pi (t) = -\partial_t K(W(q,p,t)||W_{0}(q,p)),
\end{equation}
where $W_{0}(q,p)$ is the Wigner function of the equilibrium state of the system. Here
\begin{equation}
K(W_a||W_b)=\int dq dp W_a(q,p)\ln\biggl( \dfrac{W_a(q,p)}{W_{b}(q,p)} \biggr)
\end{equation}
 is the relative Wigner entropy between the Wigner functions $W_{a,b}(q,p)$. Recently, this framework has been successful in experimentally characterizing the degree of irreversibility of the non-equilibrium dynamics of both an optomechanical system and an intra-cavity ultracold atomic system~\cite{BrunelliPRL}.

\subsection{Entropy production rate of the CSL dynamics}
Using the definition of entropy given in eq.\eqref{wigent} it is possible to get an analytical expression for both the entropy and the relative entropy of a Gaussian distribution, which will depend only on the covariance matrix.
\\
The Wigner entropy of a single-mode Gaussian state reads 
\begin{equation}
H(p) =S_2+1+\ln\pi =
%
  \ln(\pi  e \det\sqrt{V}),
\end{equation}
while the relative Wigner entropy can be computed explicitly as~\cite{adesso2012measuring}
\begin{equation}
K(p_1||p_2) =\frac{1}{2}\ln\biggl( \frac{{\rm det}V_2}{{\rm det}V_1} \biggr) +\frac{1}{2} \textit{Tr}(V_1 V^{-1}_2) 
-1,
\end{equation}
with 
$V^{-1}_{j} = \begin{pmatrix}
a_{j} & c_{j} \\
c_{j} & b_{j} 
\end{pmatrix} $~($j=1,2$).
We can use the dynamics of the entries of the covariance matrix of the system, and their equations of motion, to gather the behavior of the Wigner relative entropy $K({W(p,q,t)||W_{eq}})$ and the entropy production rate. Needless to say, the ambiguity in this case is the lack of a reference equilibrium state: the standard CSL model induces the unconstrained growth of the effective temperature of the system without reaching a stationary state. Therefore, in order to gather an intuition of the trend that the entropy production would follow, we compute the entropy production associated with target thermal states of growing variances, thus providing information on the features of both the Wigner relative entropy and $\Pi(t)$.

\begin{figure} [h!]
    \centering
    \includegraphics[width=\columnwidth]{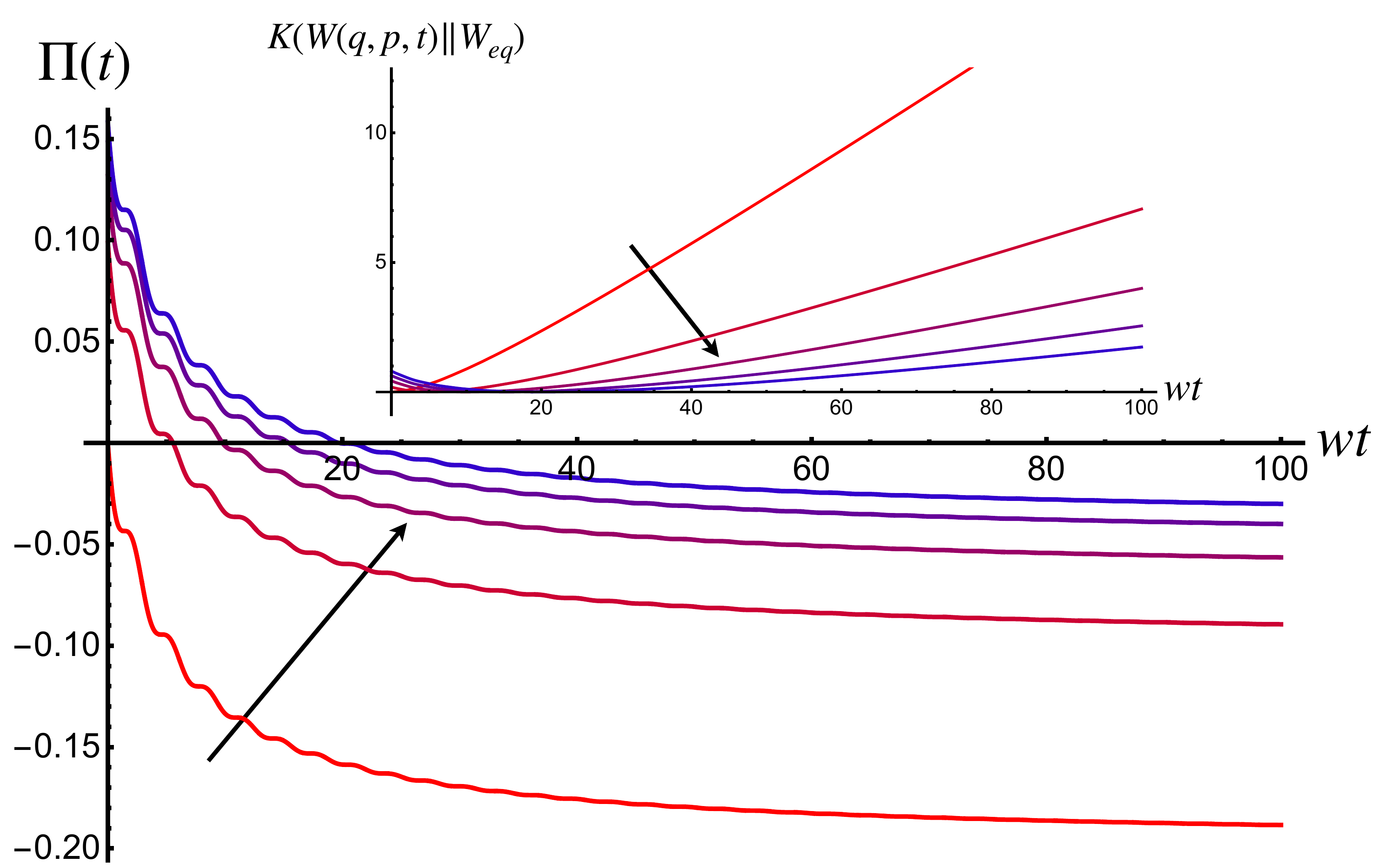}
    \caption{Entropy production rate (main panel, solid red line) and corresponding Wigner relative entropy (inset, dashed blue curve) across the dynamics. The parameters used in this simulation are $D=0.1$, ${b_1(0)}={a_1(0)}= 1/1.01$, $c_1(0) = 0$ for a target state with $a_2=b_2=1.01 k$ and $k=1,..,5$, groing in the sense of the arrows.}
\label{entropyProdMauro}
\end{figure}

\begin{figure} [t!]
    \centering
    {\bf (a)}\hskip4cm{\bf (b)}\\
    \includegraphics[width=0.5\columnwidth]{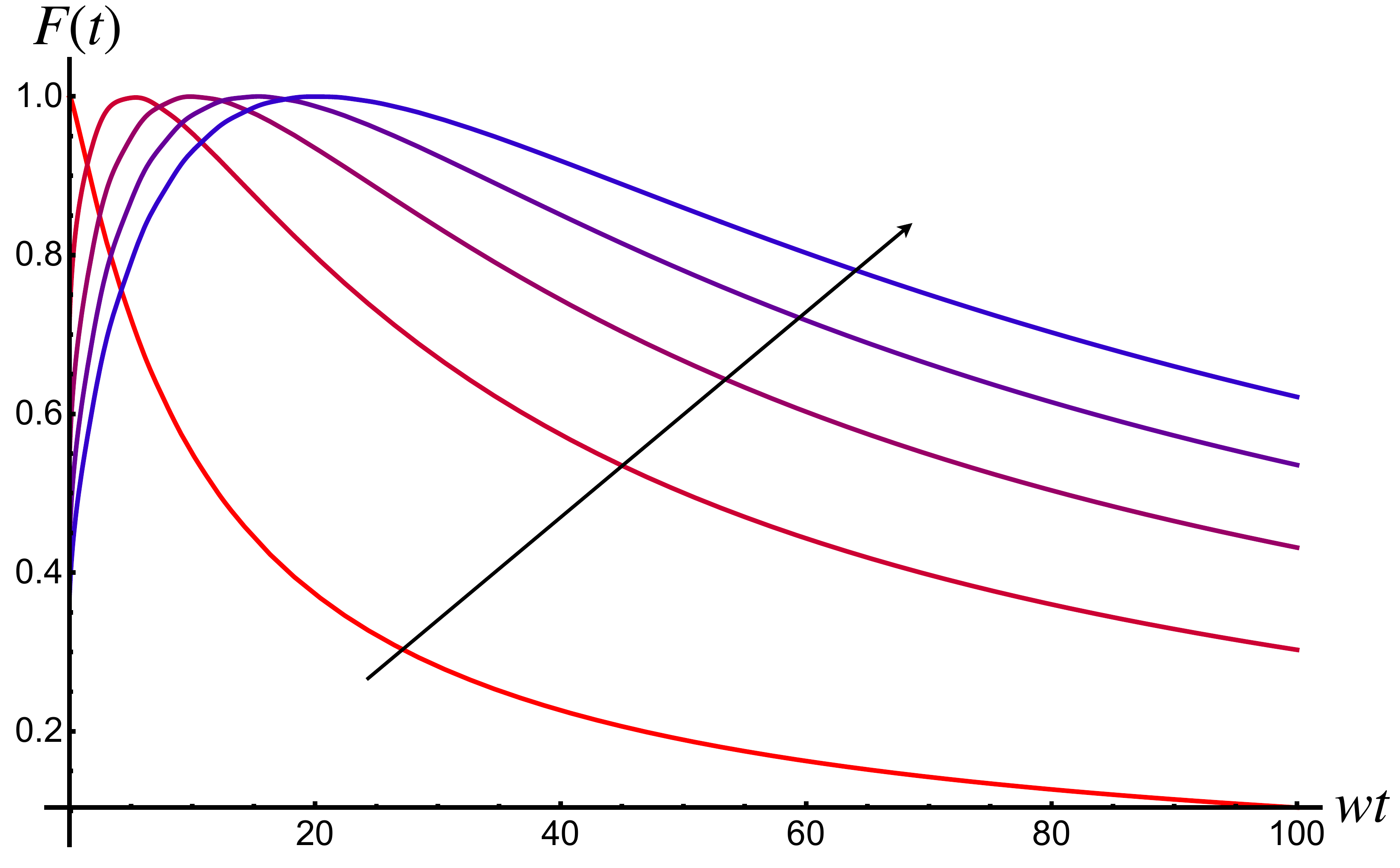}\includegraphics[width=0.5\columnwidth]{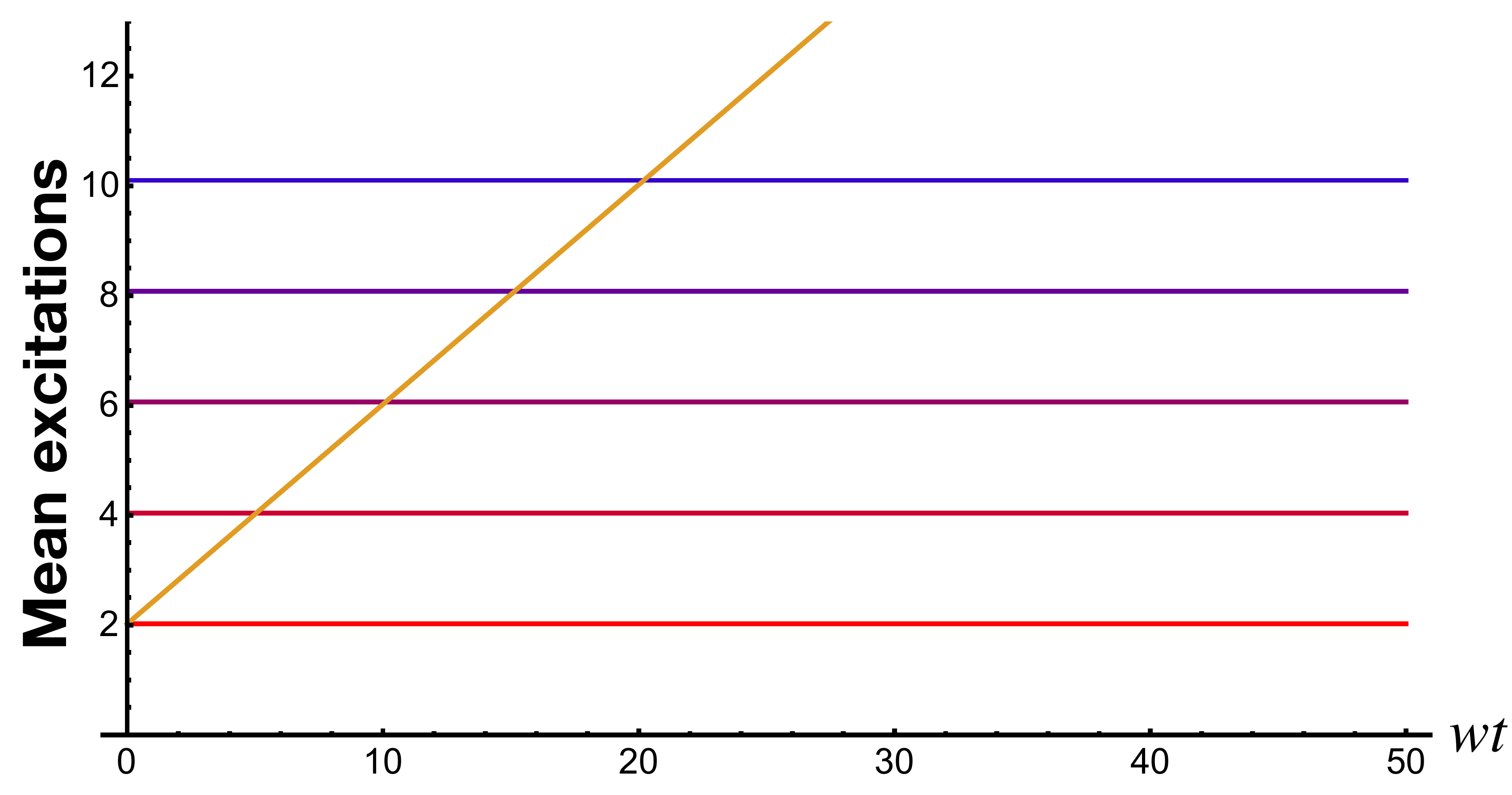}
    \caption{State fidelity [panel {\bf (a)}] and mean number of excitations in the state of the system [panel {\bf (b)}]  across the CSL dynamics. We have used the following parameters for the simulations reported int he figure: $D/(mw)=0.1$, ${1}/{b_0}={1}/{a_0}= 1.01$, $c_0 = 0$. Moreover, we have considered target states with variances $a_{eq}=b_{eq}=1.01 k$ with $k=1,..,5$ varying in steps of 1. All the reported quantities are dimensionless.}
    \label{figureFidelityCSL}
\end{figure}

Fig.~\ref{entropyProdMauro} summarizes the results of such a study. The Wigner relative entropy in general showcases a non-monotonic behavior, reaching a minimum value and then growing nearly linearly as the evolved state of the system departs from the chosen target state. Correspondingly, after remaining positive for a while, the entropy production rate takes negative values, thus witnessing the violation of the second law embodied by the constraint $\Pi(t)>0$. The minimum of relative entropy is attained at the time when the evolved state of the system becomes as close as allowed by the dynamics to the thermal state of reference. This can be clearly seen from the state fidelity between $\hat\rho(t)$ and the hypothetical reference state here at hand. Such figure of merit can be calculated straightforwardly by using the covariance matrices $\Sigma(t)$ and $\Sigma_{eq}$ of such states as~\cite{Isar2008,Nha2005,Olivares2006,Scutaru1998,Banchi}
\begin{equation}
\label{fidelity}
F(t)=\frac{2}{\sqrt{\Delta+\Lambda}-\sqrt{\Lambda}}
\end{equation}
with the symplectic invariants $\Delta=\det(V(t)+V_{eq})$ and $\Lambda=\det(\Sigma(t)+i\Omega)\det(\Sigma_{eq}+i\Omega)$, and where we have used the single-mode symplectic matrix $\Omega=\begin{pmatrix}0&1\\-1&0\end{pmatrix}$. As it can be appreciated from Fig.~\ref{figureFidelityCSL}, the state fidelity peaks at the time $wt$ when the mean number of excitations in the state of the system becomes identical to that of the target thermal state. This is also when $K(W(q,p,t)\| W_{eq})$ achieves its minimum.  %

%
Such phenomenology clearly takes place regardless of the chosen target state. This is as if the state is interacting with a thermal bath with infinite temperature: effectively the dynamics has no physical asymptotic state and thus there is no target state that could make the relative entropy disappear, reflecting the linear increase in the average energy of the system predicted by the model. 


\subsection{Entropy production rate of the dissipative CSL dynamics}

The analysis of the behavior of the entropy production rate in time can now be extended to the assessment of the dCSL mechanism, where the two dynamical regimes identified previously should be addressed separately. 


\begin{figure} [b!]
    \centering
    \includegraphics[width=0.9\columnwidth]{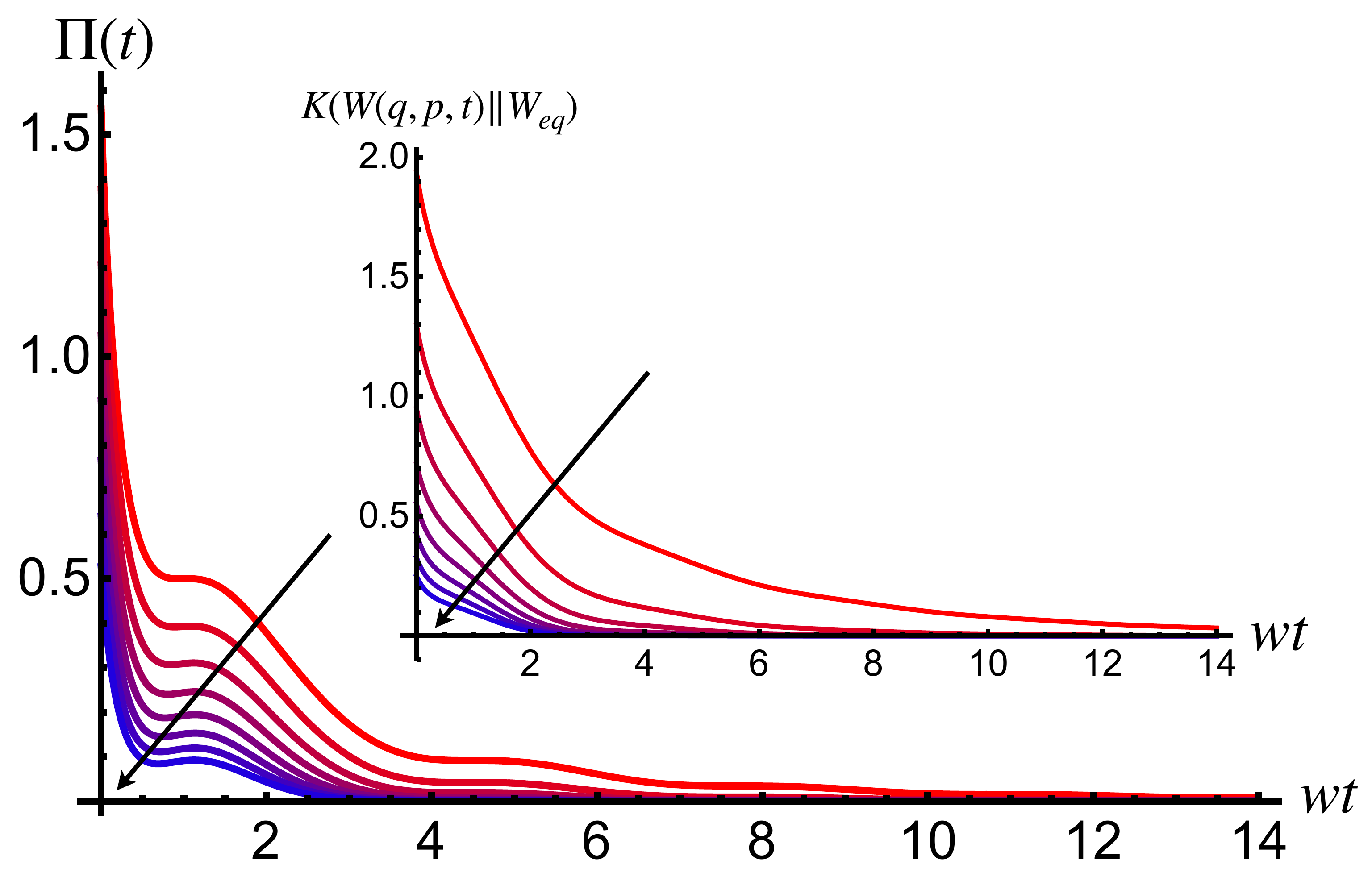}
    \caption{Entropy production rate and Wigner relative entropy (inset) over time for the dCSL model with dominant diffusion. For this simulation, we have used the parameters $D/(mw)=0.9$, and $f=0.1k$  with $k=1,..,8$, growing as shown by the sense of the arrow in the figure. We have considered initial covariance matrix elements ${1}/{b_0}={1}/{a_0}= 1.01$ and $c_0 = 0$.}
\label{FigureEntropydCSLDiffusion}
\end{figure}

\begin{figure} [t!]
    \centering
    \includegraphics[width=0.9\columnwidth]{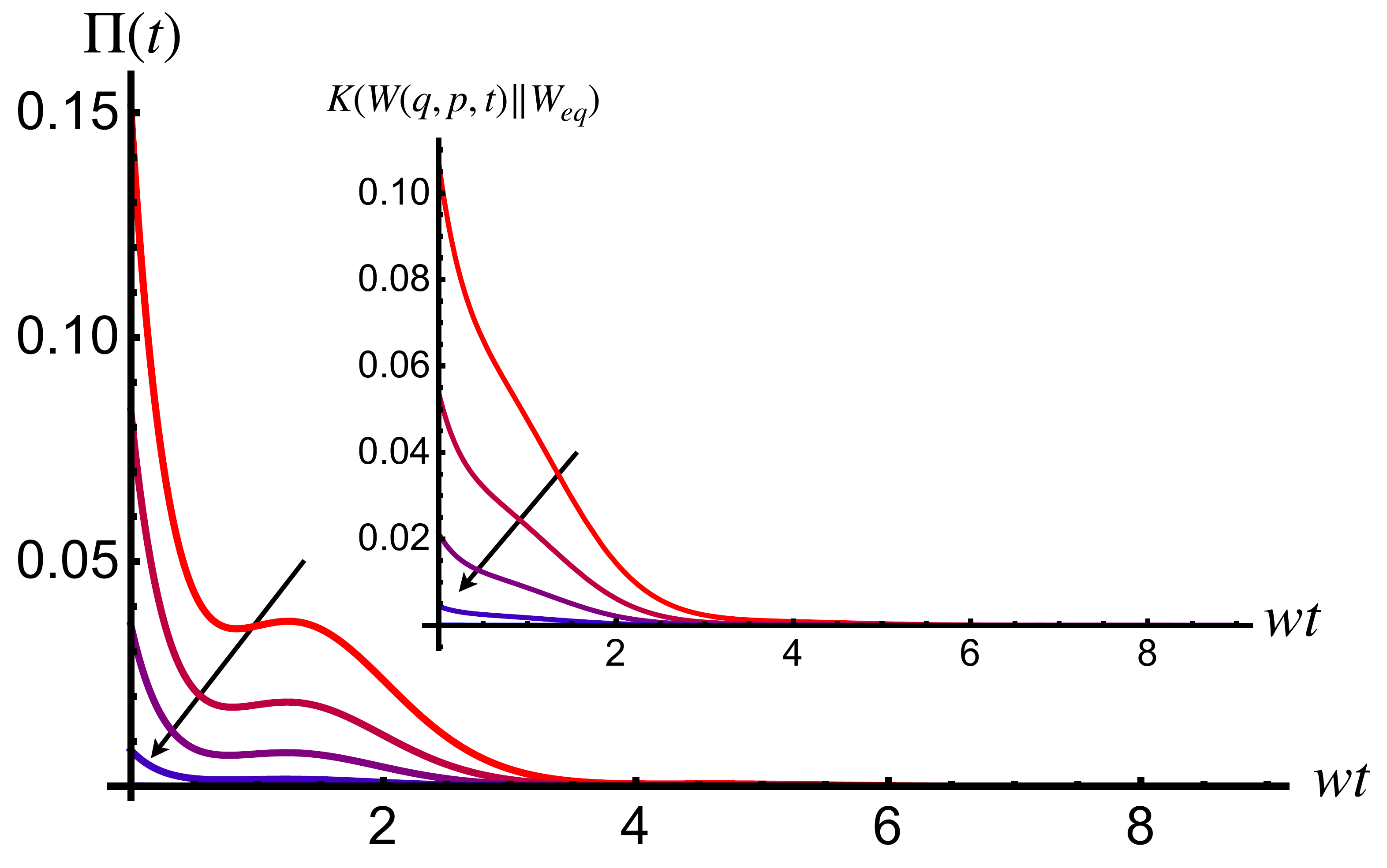}
    \caption{Entropy production rate and Wigner relative entropy (inset) over time for the dCSL model with dominant friction. {For this simulation, we have used the parameters $D/(mw)=0.5$, and $f=0.1k$  with $k=6,..,10$, growing as shown by the sense of the arrow in the figure. We have considered initial covariance matrix elements ${1}/{b_0}={1}/{a_0}= 1.01$ and $c_0 = 0$.}
    }
\label{FigureEntropydCSLFriction}
\end{figure}




  %

First, let us address the case of diffusion-dominated dynamics, where a clear stationary state is achieved as a result of the competition between diffusion and friction-like effects. The phenomenology of $\Pi(t)$ and the Wigner relative entropy is shown in Figs.~\ref{FigureEntropydCSLDiffusion} for a set of values of the parameters characterizing the dynamics. As the evolution has an asymptotic state and remains physically legitimate for any finite value of the ratio between $D/(mw)$ and $f$, after a transient, the entropy production rate $\Pi$ reduces to zero from otherwise positive values, thus satisfying the second law. 

On the other hand, care should be applied in the friction-dominated case: some values of the ratio $D/(m w f)$ may lead to physically inconsistent dynamics, as evidenced by the violations of the constraints that a legitimate covariance matrix should satisfy~\cite{Olivares2012}, namely  $V(t)\ge0$, $V(t)+i\Omega\ge0$, which implies $\vert i\Omega V(t)|\ge1$. The violation of such conditions may result in $\Pi(t)<0$ at some instant of time, thus violating the second law of thermodynamics. 
 {
A quantitative constraint comes from the uncertainty principle 
leading to $\mathrm{det}(V_{eq}) \geq 1$ or equivalently $f \leq \frac{2D}{mw}$. 

A way to obtain an inconsistent dynamics in the dCSL case is to consider a far out-of-equilibrium initial states. For instance, we have found that suitably squeezed initial states, in the friction-dominated case, might result in the violation of the second law, while still reaching an isotropic equilibrium state. Such instances do not occur, instead, for diffusion-dominated dynamics. A full characterization of the dynamics with non-isotropic initial states will be the core of a future investigation.

 

\section{Conclusions}
\label{sec:conc}
We have used a phase-space description of the dynamics entailed by both the CSL and dCSL model on a quantum harmonic oscillator, providing a thermodynamics characterisation of the dynamical features of such collapse mechanisms. Through a study of the entropy production rate, we have highlighted the lack of an equilibrium state for the case of the CSL dynamics. Correspondingly, such model violates 
the second law of thermodynamics, as showcased by a negative entropy production rate. Differently, the dynamics induced by the dCSL model indeed reaches, asymptotically, an equilibrium state for any choice of the parameters. However, the model is generally thermodynamically consistent only in the diffusion-dominated case. 

All this being said, it is clear that in general, thermodynamical transformations involving reservoirs hotter than the system are physically acceptable, only that in this case a contribution to the entropy production rate of the environment is present and must be taken into account. The present analysis is thus only partial in adressing this problem, since with a proper model of the environment that generates the noise, a non-zero temperature asymptotic state of the dynamics which does not violate the Second law could be found (with different restrictions on the parameters). A starting point could be Ref.~\cite{pearle1993ways}, where a microscopic derivation of the noise is derived. 

A similar analysis can be carried out also on other different declinations of the collapse models, such as the famous Di\'osi-Penrose model \cite{diosi2013gravity, penrose1996gravity}, which involves gravity, or energy conserving formulation of the CSL.

\acknowledgements
The authors thanks Angelo Bassi for discussions and support. 
MP acknowledges the support by the European Union's Horizon Europe EIC Pathfinder project QuCoM (Grant Agreement No.\,101046973), the Leverhulme Trust Research Project Grant UltraQuTe (grant RGP-2018-266), the Royal Society Wolfson Fellowship (RSWF/R3/183013), the UK EPSRC (EP/T028424/1), and the Department for the Economy Northern Ireland under the US-Ireland R\&D Partnership Programme.

\section*{Appendix: derivation of the Fokker-Planck equation of the CSL model}
We provide the full derivation of the quantum Fokker-Planck equation of a Gaussian state subjected to CSL used in Sec.~\ref{sec:FP}. As done previously, natural units and dimensionless position and momentum will be considered. The Weyl symbol of the statistical operator is called the Wigner function and is defined as the Fourier transform of the quantum characteristic function $\chi_{\rho}(\vec{s}) = {\rm Tr}[\hat{\rho}\hat{D}(s)]$ where $\hat{D}$ is the displacement operator \cite{navarretebenlloch2022introduction}. We have
\begin{equation*}
W_{\hat{\rho}}(\vec{r}) = \int \frac{d^2 s}{(4\pi)^2} e^{-\frac{i}{2}\vec{r}^T \Omega \vec{s}} \chi_{\hat{\rho}}(\vec{s})\> 
\end{equation*}
with $\vec{r}=(q,p)$. This expression can be shown to be equivalent to
\begin{equation*}
W_{\hat{\rho}}(\vec{r}) = \int \frac{dy}{4\pi} e^{-\frac{i}{2}py}\braket{x+\frac{y}{2}}{\hat{\rho}\bigl( x-\frac{y}{2}\bigr)} \> ,
\end{equation*}

and as this equivalence does not depend on the choice of the quantum operator which is to be transformed, it holds for any other Weyl symbol as well, that is $W_{\hat{A}}(q,p) = \int \frac{dy}{4\pi} e^{-\frac{i}{2}py}\braket{q+\frac{y}{2}}{\hat{A}\bigl(q-\frac{y}{2}\bigr)}$. Now we can use this expression to take the Weyl symbol of Eq.~\eqref{mastereqcsl}. One can start considering the time evolution of the matrix elements in the position basis of the statistical operator, derived for example in Ref.~\cite{bassi2003dynamical}
\begin{multline}
\frac{\partial}{\partial t}\braket{\vec{q'}}{\hat{\rho}(t)\vec{q''}} = -i\braket{\vec{q'}}{[\hat{H},\hat{\rho}(t)]\vec{q''}} \\
- \gamma\bigl(\frac{\alpha}{4\pi} \bigr)^{\frac{3}{2}} \bigl[ 1 - e^{-\frac{\alpha}{4}(\vec{q'}-\vec{q''})^2} \bigr] \braket{\vec{q'}}{\hat{\rho}(t)\vec{q''}} \> .
\end{multline}
Then, once specialized to the one-dimensional case with one particle, it will be enough to take the Fourier transform to get the Weyl symbol of the equation. The left-hand side will be, of course, $\partial_t W_{\hat{\rho}}(q,p)$ as the time derivative can be taken out of the integral. As for the right-hand side, let us consider first the non-Hamiltonian term, which will lead to
\begin{widetext}
\begin{equation*}
\begin{aligned}
& - \frac{\gamma}{4\pi} \sqrt{\frac{\alpha}{4\pi}}\int dy e^{-\frac{i}{2}py} \bigl[ 1 - e^{-\frac{\alpha
}{4}(q+\frac{y}{2}-q+\frac{y}{2})^{2}}\bigr]\braket{q+\frac{y}{2}}{\hat{\rho}(t)\bigl(q-\frac{y}{2}\bigl)} \\
&= -\gamma\sqrt{\frac{\alpha}{4\pi}}\biggl[ W_{\hat{\rho}}(q,p) - \frac{1}{4\pi}\int dy e^{-\frac{i}{2}py} e^{-\frac{\alpha
}{4}y^2}\braket{q+\frac{y}{2}}{\hat{\rho}(t)\bigl(q-\frac{y}{2}\bigl)} \biggr]\\
%
&= -\gamma\sqrt{\frac{\alpha}{4\pi}}\biggl[ W_{\hat{\rho}}(q,p)     - \frac{1}{4\pi} \int \frac{dk}{\sqrt{\pi}} e^{-k^2} \int dy e^{ \frac{i}{2} y (p - 2\sqrt{\alpha} k)} \braket{q+\frac{y}{2}}{\hat{\rho}(t)\bigl(q-\frac{y}{2}\bigl)} \biggr]\\
&= -\gamma\sqrt{\frac{\alpha}{4\pi}}\biggl[ W_{\hat{\rho}}(q,p)     - \int \frac{dk}{\sqrt{\pi}} e^{-k^2} W_{\hat{\rho}}(q,p-2\sqrt{\alpha} k)\biggr]\> ,
\end{aligned}
\end{equation*}
\end{widetext}
where we have used the following identity
\begin{equation*}
\int \frac{dk}{\sqrt{\pi}} \exp\bigl(-(k^2 + i\sqrt{\alpha}yk) \bigr) = \exp\biggl(-\frac{\alpha}{4}y^2 \biggr)\> .
\end{equation*}
For the Hamiltonian part we refer to Ref. \cite{manko2002alternative} for the definition of the Moyal bracket as $\bigl\{ W_{\hat{A}}, W_{\hat{B}} \bigr\}_* = -iW_{[\hat{A},\hat{B}]}$, finally getting
\begin{equation}
\begin{aligned}
&\partial_t W_{\hat{\rho}}(q,p)  =  \bigl\{ W_{\hat{H}}, W_{\hat{\rho}} \bigr\}_*(q,p)\\
& -\gamma\sqrt{\frac{\alpha}{4\pi}}\biggl[ W_{\hat{\rho}}(q,p)     - \int \frac{dk}{\sqrt{\pi}} e^{-k^2} W_{\hat{\rho}}\bigl( q,p-2\sqrt{\alpha} k \bigr)\biggr].
\end{aligned}
\end{equation}
As we are interested only in Gaussian states, we can simplify this expression through the Kramers-Moyal expansion. First, let us rearrange the integral by making a change of variable $p' = p-2\sqrt{\alpha} k$, thus getting
\begin{equation*}
 \int \frac{dk\,e^{-k^2} }{\sqrt{\pi}} W_{\hat{\rho}}\bigl( q,p-2\sqrt{\alpha} k \bigr)=- \int \frac{dp'}{\sqrt{\pi}} \frac{e^{-\frac{(p-p')^2}{4\alpha ^2}}}{2\sqrt{\alpha}}W_{\hat{\rho}}(q,p') \> .
\end{equation*} 
Considering only states whose Wigner function is well-localized around the origin (e.g. Gaussian states) one can Taylor expand the Wigner function around $p'=p$ to get an easier expression truncating at the first non trivial order, leading to
\begin{widetext}
\begin{equation*}
\begin{aligned}
\int \frac{dp'}{\sqrt{\pi}} \frac{e^{-\frac{(p-p')^2}{4\alpha ^2}}}{2\sqrt{\alpha}}W_{\hat{\rho}}(q,p')\simeq&\int dp' \frac{e^{-\frac{(p-p')^2}{4\alpha}}}{\sqrt{4\pi \alpha}} \bigl[ W_{\hat{\rho}}(q,p) + \partial_p W_{\hat{\rho}}(q,p)(p'-p)+ \partial^2_{p}W_{\hat{\rho}}(q,p)(p'-p)^2 \bigr] 
\end{aligned}
\end{equation*}
\end{widetext}
where the approximation sign is used as higher order terms have been neglected. 
As the second term is the integral is identically zero in light of parity, after some algebra one is left with
\begin{equation*}
W_{\hat{\rho}}(q,p) + 2\alpha \partial^2_{p}W_{\hat{\rho}}(q,p)\> .
\end{equation*}
Thus the Fokker-Planck equation will be of the form
\begin{equation*}
\partial_t W_{\hat{\rho}}(q,p) = \bigl\{ W_{\hat{H}}, W_{\hat{\rho}} \bigr\}_*(q,p) +\sqrt{\biggl(\frac{\gamma ^2 \alpha ^3}{\pi}\biggr)}\partial^2_{p}W_{\hat{\rho}}(q,p)\> .
\end{equation*}

\bibliography{biblio.bib}

\end{document}